\documentclass[runningheads]{llncs}
\usepackage{graphicx}
\usepackage{tabularx}
\usepackage{multirow}

\label{a: document}
\begin{document}

\label{a: title}
\title{A Characterisation of Smart Grid DoS Attacks}
%

\label{a: authors}
\author{Dilara Acarali\inst{1} \and
        Muttukrishnan Rajarajan\inst{1} \and
        Doron Chema\inst{2} \and
        Mark Ginzburg\inst{2}}

\authorrunning{D. Acarali et al.}

\institute{School of Mathematics, Computer Science \& Engineering. City, University of London. London, UK.\\
\email{\{dilara.acarali.2, r.muttukrishnan\}@city.ac.uk}\\
\url{www.city.ac.uk/about/schools/mathematics-computer-science-engineering} \and
Technical Team, L7 Defense, BeerSheva, Israel\\
\email{\{doron, marik\}@l7defense.com}\\
\url{www.l7defense.com/}}

\maketitle 

\begin{abstract}

Traditional power grids are evolving to keep pace with the demands of the modern age. Smart grids contain integrated IT systems for better management and efficiency, but in doing so, also inherit a plethora of cyber-security threats and vulnerabilities. Denial-of-Service (DoS) is one such threat. At the same time, the smart grid has particular characteristics (e.g. minimal delay tolerance), which can influence the nature of threats and so require special consideration. In this paper, we identify a set of possible smart grid-specific DoS scenarios based on current research, and analyse them in the context of the grid components they target. Based on this, we propose a novel target-based classification scheme and further characterise each scenario by qualitatively exploring it in the context of the underlying grid infrastructure. This culminates in a smart grid-centric analysis of the threat to reveal the nature of DoS in this environment.

\end{abstract}

\section{Introduction}

The digital age has caused an increased dependency on electricity, and consequently, given rise to an increased demand on power systems. As a result, traditional power grids have had to evolve to deal with this inflated demand. A smart grid is a traditional power grid integrated with information communication systems. The former is referred to as the physical or operational technology (OT), and the latter is called the cyber or information technology (IT). In practice, this means that IT networks gather data from field systems and deliver them to a central command centre via local controllers. That data can then be used to regulate physical grid components and to make management decisions.

Whilst this can greatly improve efficiency, the IT network also makes the smart grid more vulnerable to malicious activity by expanding the previously limited attack surface. Grid components are now remotely accessible, and grid processes are now dependent on data flows through communication channels that can be disrupted. Furthermore, the cyber and the physical systems are highly interconnected and interdependent, meaning that faults or attacks at one point can cause a chain of effect across the wider smart grid.

This work is focused on Denial-of-Service (DoS), a well-known cyber-attack targeting availability, designed to hinder normal system processes. It is popular because, despite being relatively simple, a successful DoS attack can cause a large degree of disruption. DoS attack methodology may consist of a) flooding, where a channel/device is overwhelmed with data, b) the exploitation of vulnerabilities or quirks in systems and protocols, or c) both. A DoS attack launched by multiple dispersed individuals (e.g. in a botnet) is known as Distributed DoS (DDoS). Whilst disruption resulting from physical tampering can also be explored, we consider DoS predominantly as a cyber threat.

DoS attacks in conventional networks are well-studied, but the smart grid has particularities that influence both methodology and results. In this paper, we first identify and then characterise smart grid DoS scenarios to build up a picture of how this threat manifests in this new environment. Identification is achieved via a detailed survey of existing research, which then forms the basis of a new classification scheme organised by potential targets. This differs from conventional approaches and is designed to link targeted grid components with likely DoS attack methods, providing a reference for researchers and defenders. To our knowledge, a survey and classification of smart grid-specific DoS scenarios is novel to this work.

The research presented in this paper is part of the European Union's \textit{Energy Shield} project \cite{EnergyShield}, commissioned in recognition of the energy industry's transition from traditional systems to smart grids. The aim of this project is to develop a defence toolkit for EPES (Electrical Power and Energy System) operators \cite{EnergyShield} to protect critical infrastructure from cyber-attacks, including DDoS. This work was conducted to provide a foundation of understanding of the smart grid-centric DoS threat on which the project can further build.

The rest of the paper is structured as follows: Section 2 provides a background explanation of smart grid architecture, including domains, flows, and key components. Then, the smart grid DoS research survey is presented in Section 3. Based on the information in 2 and 3, Section 4 presents our classification scheme and characterisations of the identified DoS scenarios. This is followed by the discussion of our findings in Section 5, with an analysis of related works given in Section 6. We then conclude in Section 7.

\begin{figure}[!t]
\label{fig1: domains}
\centering
\includegraphics[scale=0.27]{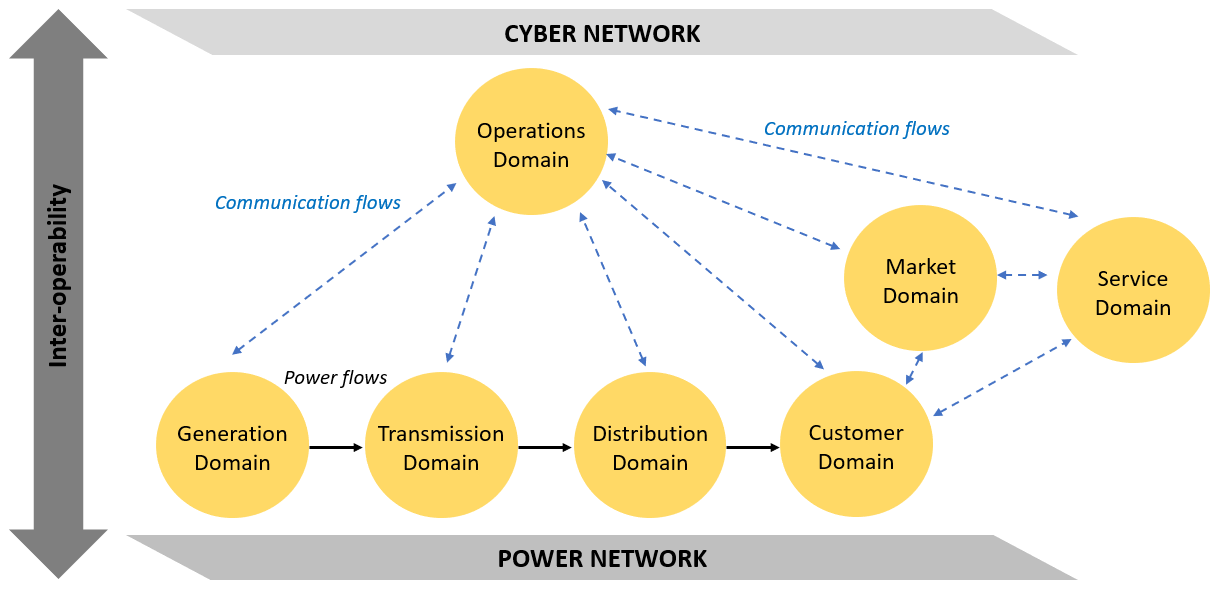}
\caption{Smart grid domains.}
\end{figure}

\section{Smart Grids Background}

\subsection{Smart Grid Domains}

The smart grid follows a domain model architecture, which means it is split into multiple domains, each handling a different function. A domain will generate/process either power flows (in the OT), communication flows (in the IT), or both. Generally, data on power flows is collected and shared as communication flows. Fig. 1 provides an illustration of the core domains as they relate to each other and the flows between them. Note that this paper focuses on communication flows as the main target of DoS attacks, but the characteristics of these are influenced by the nature of the underlying power flows.

The core domains of the traditional power grid, responsible for the production and delivery of electricity as a utility, are described first. These are the ones that contain power flows. The Customer domain deals with the delivery of power to customer premises. It also handles the collection of usage data. Power is received from the Distribution domain, which is responsible for dissemination. This domain also transforms the power received from the Transmission domain, which is where bulk energy is carried between geographically distributed locations (e.g. a power plant and a city). The Generation domain houses power plants where electricity is `produced'. If distributed power generation exists (i.e. customers generate their own electricity), this is also handled within the Customer and Distribution domains.


The rest are the communication-focused domains. Operations is the main control hub and the core of the communication network, responsible for the collection of monitoring data and the dissemination of control commands. The Service domain houses the providers who deliver electricity as a utility, whilst the buying/selling of electricity is handled within the Markets domain. Both Service and Markets make use of the IT to provision services and to bill customers. However, it should be noted that there is a separation between the IT that controls the grid, and the corporate TCP/IP networks of energy companies and service providers.

\subsection{Smart Grid Structure}

The traditional power grid (i.e. the OT) is hierarchical in the way it transmits and distributes electricity to users. One or more power plants generate electricity. This is then carried in bulk through heavy-duty transmission lines to many geographically dispersed substations. Here, transformers convert (i.e. step-down) and transfer energy to distribution lines, which then branch out to deliver power to a large number of individual customer premises. Hence, there are a larger number of component systems at the bottom then at the top, which means that a single problem higher up in the power hierarchy affects many systems lower down.

The IT network is also hierarchical. For a given region and/or provider, a single core management system, such as SCADA (Supervisor Control and Data Acquisition), contains various master controls to monitor, analyse, and regulate grid operations. This core network receives inputs from distributed monitoring devices (like RTUs and PLCs, described in the next sub-section) sitting in FANs (Field Area Networks), NANs (Neighbourhood Area Network) and substation LANs (Local Area Networks) operating throughout the Generation, Transmission, and Distribution domains. NANs amalgamate smaller customer networks, including HANs (Home Area Networks), BANs (Business Area Networks), and IANs (Industry Area Networks). Each contain devices that connect to upload usage data. As with the OT, the IT hierarchy means that higher level issues impact a large number of lower level systems. Furthermore, there is a variation in DoS attack surface (smaller at the top, bigger at the bottom).

\begin{figure}[!t]
\label{fig2: IT heirarchy}
\centering
\includegraphics[scale=0.30]{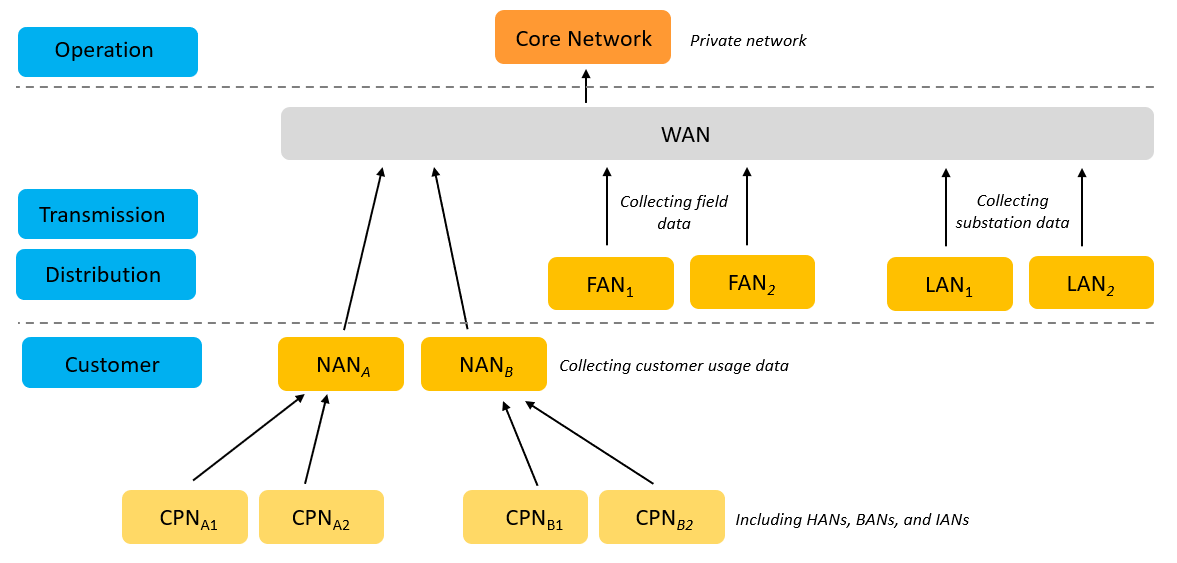}
\caption{Smart grid IT network hierarchy with CPNs (Customer Premise Networks).}
\end{figure}

In addition to this, connectivity and dependency exist between the IT and OT, which means that malicious activity in one will have some influence on the other. In other words, assuming that DoS attacks are targeted at IT devices/flows, when some function of the IT network is denied, there will be a corresponding impact, related to that function, on the OT. Furthermore, disruption to a particular grid process can have subsequent effects on other processes, and may escalate into general grid instability. This is a unique characteristic of critical infrastructures, as DoS within conventional IT networks does not typically have the potential for this level of widespread disruption.

\subsection{Smart Grid Components}

\begin{table}[t!]
\label{table1: grid devices}
\centering
\begin{tabularx}{\textwidth}{l|l|X}
  \textbf{Systems} &\textbf{Domains} &\textbf{Function}\\
  \hline
  Smart Meter        &-Customer     &Located in customer premises, collects usage data for operations (via the AMI) and provides pricing information to users. \\ \hline

  Remote Telemetry   &-Distribution &\multirow{3}{\hsize}{Located in substations, collects data on grid processes from PLCs for automated monitoring and control processes.} \\
  Unit (RTU)         &-Transmission & \\
                     &-Generation   & \\ \hline

  Phasor Measurement &-Distribution &\multirow{3}{\hsize}{Located in substations, collects data on electrical phasor patterns for synchronisation of grid supply and demand.} \\
  Unit (PMU)         &-Transmission & \\
                     &-Generation   & \\ \hline

  Programmable Logic &-Distribution &\multirow{2}{\hsize}{Field device, collects telemetry data on grid processes, communicated to RTUs and LFC to generate control signals.} \\
  Controller (PLC)   &-Transmission & \\
                     &-Operations   & \\ \hline

  Master Telemetry   &-Operations   &\multirow{2}{\hsize}{Part of SCADA or WAMS, processes data received and aggregated from RTUs.} \\
  Unit (MTU)         &              & \\ \hline

  Phasor Data        &-Distribution &\multirow{2}{\hsize}{Part of SCADA or WAMS, processes data received and aggregated from PMUS.} \\
  Concentrator (PDC) &-Transmission & \\
                     &-Generation   & \\
                     &-Operations   & \\ \hline

  Load Frequency     &-Generation   &\multirow{2}{\hsize}{Located alongside generators, minimises fluctuations in energy input and output for frequency balance.} \\
  Control (LFC)      &              & \\
                     &              & \\ \hline

  Supervisory Control&-Operations   &\multirow{3}{\hsize}{Control system, responsible for manipulating grid topology, monitoring processes, and maintaining functionality.} \\
  \& Data Acquisition&              & \\
  (SCADA)            &              & \\ \hline

  Wide Area          &-Operations   &\multirow{3}{\hsize}{Control system, uses information gained from PMU/PDC data to monitor and react to grid instabilities/issues.} \\
  Management System  &              & \\
  (WAMS)             &              & \\ \hline
\end{tabularx}
\caption{Smart grid sub-systems and components, their domains, and their functions.}
\end{table}

Within the IT network, there are a number of sub-systems responsible for different grid processes. The key sub-systems are:

\noindent \textbf{-Advanced Metering Infrastructure (AMI)}: A two-way communication network between smart meters sitting within customer premises and utility servers in the core network. The AMI enables the collection of usage data (analysed for load forecasting and pricing models), as well as the delivery of relevant customer services.

\noindent \textbf{-Phasor Control}: Consists of PMUs (Phasor Measurement Units) which measure electrical signals and monitor phasor patterns, and PDCs (Phasor Data Concentrators) which aggregate and process this data for monitoring, fault response, and command generation. The purpose of this is to ensure that the frequency of the electricity in the grid is synchronised and the grid remains stable.

\noindent \textbf{-Telemetry Control}: Consists of RTUs (Remote Telemetry Units, distributed across the domains), which collect telemetry data from grid components, and MTUs (Master Telemetry Units, connected to core systems), which receive that data and process it for management and topology manipulation. This supports efficient power generation and transfer. RTUs connect multiple PLCs (Programmable Logic Controllers), which connect to field devices.

\noindent \textbf{-Load Frequency Control}: Consists of PLCs (that connect to field devices) and RTUs which collect data on the performance of various processes. This is communicated to LFCs (Load Frequency Controls), which use the data to manage generators to maintain a stable frequency in the grid.

\noindent \textbf{-Core Control Systems}: Includes SCADA (Supervisor Control and Data Acquisition) systems and WAMS (Wide Area Management Systems), which act as the central management point for the grid, where all data is aggregated, processed, and used for human-lead decision-making and for automatically generated controls.

Table 1 provides a summary of typical sub-systems and components, which domains they sit in, and what they do. Note that this list is not exhaustive. For the scope of this paper, we have focused on the most common sub-systems which rely on communication flows. Other systems include those responsible for pricing models, for distributed power generation management, and external data systems used for load prediction (e.g. weather forecasts).

\section{Smart Grid DoS Survey}

The purpose of this survey is to answer the following research question: \textit{According to the literature, what are the DoS attack possibilities in smart grids, given the smart grid's unique characteristics?} The answer to this question must consider attack method, attack target, and attack impact. The survey methodology used is as outlined in \cite{brereton2007}, and the review protocol was to search IEEEXplore, Science Direct, and Google Scholar with a set of DoS and smart grid search term pairs. Specifically, we aimed to identify recent works that defined, modelled, simulated, or discussed DoS scenarios. Note that SCADA or WAMS are not considered, as these systems are common across critical infrastructures and are considered a separate field of study. The scenarios identified in the surveyed works are summarised here.

Wang et al. (2017) \cite{wang2017} explored the adversarial interaction between smart grid defenders and attackers, anchored on an AMI DDoS attack. The AMI is modelled as a tree with many smart meters connecting to aggregators in layers. Traffic from the meters travels up these layers to a base station, which then relays it to the AMI core. In this study, the attacks were targeted at smart meters and aggregators, assuming the attack source to be a botnet. The authors found that, depending on where the target sits, certain communication paths become saturated, with nodes attached to these paths consequently being knocked offline. Meanwhile, the AMI's tree structure eventually causes ``downstream'' nodes to lose core connectivity \cite{wang2017}. Honeypots embedded within the AMI core were used to derive optimal attack and defence strategies.

As with \cite{wang2017}, Guo and Ten (2015) \cite{guo2015} also studied a botnet-driven AMI DDoS scenario. They created a two-staged model combining botnet formation and attack launch. Three actor categories were considered: attackers, victims, and agents (i.e. smart meters converted into bots). The authors posited that it is reasonable to assume that a population of smart meters will have similar vulnerabilities, and hence, will be susceptible to the spread of an automated malware targeting firmware or communication functions \cite{guo2015}. They simulated a UDP flood against a 2-layer AMI topology, with each bot generating packets at a rate of 2Mbps \cite{guo2015} and reported that growth in the bot population directly correlated with an increased number of dropped packets and longer end-to-end delays \cite{guo2015}.

Similarly, Sgouras et al. (2017) \cite{sgouras2017} investigated the AMI impact of a botnet-launched DDoS attack. They modelled the AMI as multiple residential smart meters connected to a central control server, and posited that the Internet is a likely channel for communication between control servers and aggregators. Based on this, they suggested that a botmaster could sniff traffic to determine the server IP and then use a botnet (external to the smart grid network) to launch a TCP SYN DDoS attack at great scale. Using this proof-of-concept, the authors were able to demonstrate how Internet-connectivity exposes the grid to the outside world and can make it susceptible to attacks from remote adversaries.

Asri and Pranggono (2015) \cite{asri2015} also investigated botnet-based DDoS attacks against the AMI. With similar assumptions to \cite{sgouras2017}, they modelled the AMI as a collection of households containing smart meters connecting to a central utility server via the Internet. An external botnet can then connect to the utility server to flood it in a UDP storm attack, targeting many random ports. The server will try to initiate applications on random ports that do not exist and respond with ICMP `Destination Unreachable' packets \cite{asri2015}. With simulations, the authors showed that the entire grid could be compromised with a large-enough DDoS attack, though the effect on the power supply network was not immediate. Only after the server had been knocked offline was an impact observed.

Sgouras et al. (2014) \cite{sgouras2014} considered four different AMI DoS setups: 1) DoS on a smart meter, 2) DDoS on a smart meter, 3) DoS against an AMI utility server, and 4) DDoS on an AMI utility server. Comparing the DoS and DDoS scenarios, they found that the targeted smart meter suffered from significantly increased queue lengths under the latter. In fact, they observed that queue lengths reached maximum levels much faster. The DoS attack against the server caused a drop in the number of TCP packets delivered to smart meters, leading to some service degradation. In comparison, the DDoS attack on the server reportedly diminished connections with almost 90\% of the smart meters \cite{sgouras2014}.

Hoffman and Bumiller (2019) \cite{hoffmann2019} proposed a special AMI DoS attack called Denial-of-Sleep. This is where a battery-powered device is prevented from entering sleep mode (i.e. a low-power state intended to conserve energy), thereby significantly reducing its lifespan. Smart meters enter sleep mode when they are not forwarding measurement data or receiving traffic from other nodes \cite{hoffmann2019}. Two sleep protocols are identified: S-/T-mode (where the device transmits and waits some time for a response before sleeping) and C-mode (where the device sleeps immediately after completing its transmission). The authors used an abstract version of a TLS (Transport Layer Security) handshake to model Denial-of-Sleep attacks in the context of the C-mode sleep protocol and reported that a small number of attacks of relatively short length can significantly deplete batteries \cite{hoffmann2019}.

Chatfield et al. (2018) \cite{chatfield2018} studied jamming, which they categorised as a form of DoS used to disrupt the wireless networks within smart grids. They defined two possible scenarios. The first is where a jamming attack produces lots of radio signals on the same frequency as legitimate communications, causing delays and increased latency for control messages. The second is where the degradation caused by a jamming attack interferes with standard protocol processes and leads to constant retransmissions, further congesting the network and, in the case of routing protocols, causing network instability \cite{chatfield2018}. In their AMI model, they used the received signal strength of nodes to differentiate between normal and attack scenarios, and related attack effectiveness to the distance between attacker and victim nodes.

Pedramnia and Rahmani (2018) \cite{pedramnia2018} explored AMI-based DoS against cellular LTE (Long-Term Evolution) networks. They identified signalling attacks, where bearer assignment is exploited to prevent legitimate use. A malicious bearer request is sent, and once an assignment is made, that bearer is left unused. It expires, triggering another assignment process, and the pattern is repeated. LTE-specific jamming attacks may be possible too. The authors also discussed SMS link saturation, where device user panels (where customers receive updates) are flooded. This drowns out legitimate messages, causes buffer overflows, or leads to delays \cite{pedramnia2018}. Lastly, DoS against NAT (Network Address Translation) systems are highlighted, specifically for NAT64. These include NAT overflow (where malicious mapping requests block legitimate use), NAT wiping (where TCP-RST messages are used to delete mappings), and NAT breaking (where spoofed IPs make NAT requests and ignore server responses, forcing those IPs onto blacklists) \cite{pedramnia2018}.

\begin{table}[t!]
\label{table2: survey}
\centering
\begin{tabularx}{\textwidth}{l|X|l|X}
  \textbf{Ref}        &\textbf{Premise}                    &\textbf{Target} &\textbf{Impact}\\ \hline

  \cite{wang2017}     &Defender vs. attacker interactions. &AMI &Channel saturation; downstream nodes lose core connectivity. \\ \hline

  \cite{guo2015}      &Botnet-launched UDP flood DDoS. &AMI &Botnet growth increases dropped packets and end-to-end delays. \\ \hline

  \cite{sgouras2017}  &Botnet-launched TCP-SYN DDoS. &AMI &Load fluctuations and system availability diminished. \\ \hline

  \cite{asri2015}     &UDP storm DDoS against residential AMI. &AMI &Possible complete compromise; delayed impact on power network. \\ \hline

  \cite{sgouras2014}  &DoS/DDoS attack impact in the AMI. &AMI &Dropped TCP packets, service degradation, and diminished connectivity. \\ \hline

  \cite{hoffmann2019} &Denial-of-Sleep attacks against battery-powered AMI nodes. &AMI &Increased battery depletion in affected nodes. \\ \hline

  \cite{chatfield2018} &Jamming attack detection in wireless networks. &AMI &Delays on legit frequencies; protocol process interruption.\\ \hline

  \cite{pedramnia2018} &Signalling, SMS link saturation, and NAT attacks against cellular LTE networks. &AMI &Repeated bearer assignments; reduces devices' interface functionality. NAT-based disruptions. \\ \hline

  \cite{yi2016}        &Puppet attack against mesh networks. &AMI &Corrupted address lists, causing path discovery cycles. \\ \hline

  \cite{wei2012}       &Resilient routing model for PMUs-WAMS traffic. &PMUs &Delays/packet dropping within PMU data channels. \\ \hline

  \cite{srikantha2015} &Nash Equilibrium-based DoS-resilient routing. &PMUs &Reduced relay nodes functionality; delays/reduced connectivity.\\ \hline

  \cite{yilmaz2018}    &Interface and PLC-targeting attacks exploiting query replies. &PLCs &Delays increase with attack length, management SW becomes non-functional.\\ \hline

  \cite{liu2013}       &DoS switching strategies against LFCs. &LFCs &Attack impact is maximised via start time selection and attack length. \\ \hline

\end{tabularx}
\caption{DoS types and targets identified by surveyed works.}
\end{table}

Yi et al. (2016) \cite{yi2016} defined a new AMI DoS technique called the puppet attack. This seeks to exploit the use of DSR (Dynamic Source Routing) and the way that mesh networks are formed. One or more nodes are selected as `puppets', and receive attack commands. DSR uses route requests (RREQ) and route replies (RREP) to build address lists amongst nodes. The attack makes a puppet node erase addresses from the list, causing path errors \cite{yi2016}. This then triggers another round of path discovery and list building, thereby stopping the network from settling into a routing structure. This contrasts with standard DoS which relies on crafted packets or exhaustion of resources. Meanwhile, puppet attacks undermine the structure and functionality of the mesh itself.

Wei and Kundar (2012) \cite{wei2012} explored DoS attacks targeting communication channels between distributed PMUs and the WAMS. Specifically, the authors suggested that PMU data rates may vary with network congestion. They modelled a hierarchical network covering the cyber (IT) and physical (OT) networks. PMUs in the cyber network collect data on a particular generator node in the physical network. Local controllers obtain data from PMUs to create control signals to be applied to the generator nodes. Generators are grouped into clusters, with a single PMU and local controller in charge of each cluster. Hence, power flow depends on control signals, and control signals depend on PMU data. The DoS attack then targets the PMUs' communication channels with the aim of causing delays or packet drops. The authors used this attack model to propose a flocking-based scheme to route traffic around DoS-affected regions.

Srikantha and Kundur (2015) \cite{srikantha2015} also studied attacks on PMUs to enhance resiliency against DoS attacks. They modelled the smart grid as a pair of hierarchical, inter-connected directed graphs, populated with relay nodes (RNs) responsible for transmitting data. Some function as PMUs (collecting data) and some as cyber-actuators (sending control signals). At the top of the hierarchy is the root node that transmits control data downstream, whilst PMUs send measurement data upstream. The authors then experimented with DoS attacks that target one or more RNs in the tree. This causes delays and disruption for the downstream/upstream movement of traffic as RNs are rendered non-functional. As with \cite{wei2012}, they proposed a routing system to allow the topology to morph around the attacked nodes.

Yilmaz et al. (2018) \cite{yilmaz2018} explored the possibility of DoS attacks against PLCs, suggesting that a PLC can be targeted both from within and outside of its own IP network, as long as its IP address is known. Furthermore, the authors highlighted that PLCs reply to any queries from any source, further increasing their potential for exploitation \cite{yilmaz2018}. A testbed was built, consisting of PLC devices and some PCs running a) TIA (Totally Integrated Automation) portal management software, b) DoS tools, and c) attack detection systems \cite{yilmaz2018}. Attacks were then simulated against both the PLCs and the TIA portal. The results showed that the PLCs' ping response delay continued to increase the longer the attack was sustained. Meanwhile, the TIA portal became non-functional. The authors noted that the network was quickly disrupted even with a small number of attackers \cite{yilmaz2018}.

Finally, Liu et al. (2013) \cite{liu2013} investigated DoS attacks designed to disrupt the delivery of telemetry data from RTUs to LFC systems. This would prevent the LFC from generating accurate command signals for physical grid components, potentially causing further issues. The authors modelled DoS as a switched system and suggested that attacks can have maximum impact if attackers select the optimal switching strategy. They identified this to be a sequential attack over multiple intervals \cite{liu2013}. DoS attacks were simulated with different starting times, revealing that impact was more significant for those launched before power systems have fully converged. This period was therefore highlighted as one of increased vulnerability \cite{liu2013}.

The survey is summarised in Table 2 and the findings are analysed in the next section.

\section{Smart Grid Denial-of-Service Characterisation}

Based on the architecture of smart grids (discussed in Section 2) and the literature survey (presented in Section 3), we propose a set of smart grid DoS scenarios, classified in a target-based structure. This contrasts with existing classification schemes which tend to focus on attack methodology. The main categories are:
\begin{itemize}
  \item \textbf{\textit{A}: Network-Targeting}
  \begin{itemize}
    \item \textbf{\textit{A.1}}: Saturation Scenarios (aiming to use up channel resources)
    \item \textbf{\textit{A.2}}: Exploit Scenarios (aiming to manipulate standard processes)
  \end{itemize}
  \item \textbf{\textit{B}: Device-Targeting}
  \begin{itemize}
    \item \textbf{\textit{B.1}}: Exhaustion Scenarios (aiming to use up device resources)
    \item \textbf{\textit{B.2}}: Compromise Scenarios (aiming to manipulate a device)
  \end{itemize}
\end{itemize}


In \textit{A} scenarios, the aim is to disrupt the network itself, either by blocking communications through full consumption of channel capacity (\textit{A.1}), or the blocking of standard operations by preventing normal protocols from functioning as intended (\textit{A.2}). Meanwhile, the aim of \textit{B} scenarios is to disrupt the operation of particular network nodes so that they can be manipulated and cannot function as normal. This may be achieved by overwhelming the capacity of a device (\textit{B.1}) or by exploiting some vulnerability in it (\textit{B.2}). Note that \textit{B.1} scenarios are similar in concept to \textit{A.1}. In the following, each scenario is described in terms of the smart grid sub-systems that may be targeted, allowing us to consider subsequent impact possibilities given the smart grid's multifaceted and inter-connected infrastructure. Key points for effective defence and mitigation are also highlighted.

\subsection{Network Saturation Scenarios (\textit{A.1})}

The AMI may use a number of communication technologies, including WiFi, WSNs, cellular networks, or the Internet. Hence, it can be targeted in several different ways. Saturation may be attempted using typical flooding attacks (e.g. ICMP flood, UDP flood, HTTP flood) on any layer of the TCP/IP protocol stack, as seen in conventional networks. This type of scenario was explored by \cite{guo2015} and \cite{asri2015}. Jamming attacks may also be used to similar effect against wireless channels, as cited by \cite{chatfield2018}. Similarly, where SMS communication is used to push information to device interfaces, SMS link saturation may be employed \cite{pedramnia2018}. Therefore, \textit{A.1} scenarios in the AMI can be characterised by a reduced upstream flow of usage data and downstream flow of service data. This could consequently lead to load estimation errors, bad pricing models, and reduced service quality for customers. Furthermore, as modelled in \cite{wang2017}, the hierarchical structure of the AMI can cause a larger number of nodes lower down in the chain to lose connectivity to the core, further exasperating the issue.

Meanwhile, the Load Frequency Control sub-system may use SCADA protocols like ICCP (Inter Control Center Protocol) \cite{knapp2011} or the IEC 61850 protocol stack \cite{mattioli2015} running on top of TCP/IP infrastructure \cite{knapp2011} \cite{elgargouri2015} to ensure RTU-to-core communication. Flooding-style attacks can therefore be deployed here too, as examined by \cite{liu2013}. \textit{A.1} scenarios in the LFC system would be characterised by the untimely or reduced sharing of telemetry data by RTUs. This could result in incorrect control signal generation and consequently, the incorrect operation of physical grid devices, which may escalate into grid instability. Furthermore, it should be noted that ICCP and IEC 61850 do not provide robust and secure authentication mechanisms \cite{knapp2011} \cite{elgargouri2015}, leaving these communications vulnerable to malicious influence. The mechanisms for how flooding may be achieved on channels using these protocols (and for DoS in general) is an area for further study.

The survey also threw up the threat of \textit{A.1} scenarios against the Phasor Control sub-system, which may use IEC 61850 protocols too \cite{mattioli2015}. This attack possibility was explored by \cite{wei2012}, where the authors examined the relationship between PMU data and control signals. The results of their experiments suggest that the prevention of timely measurement readings due to flooded communication channels can result in incorrect control signals, which in turn may lead to fluctuations in frequency and ultimately, an unstable grid. This assessment on the impact of disrupted PMU flows was also supported by the results of \cite{srikantha2015}. A similar angle may be considered for the Telemetry Control and Core Control sub-systems as well.

Saturation scenarios on TCP/IP connections may be dealt with using anomaly detection to identify increases in traffic volume. This is also applicable against low-rate DoS attacks \cite{xiang2011}. Similarly, IDS (Intrusion Detection Systems) should be used to identify suspicious activity, including jamming attacks \cite{chatfield2018}. For external attack sources, traffic from suspected IPs can then be blocked. Both anomaly detection and IDS (Intrusion Detection Systems) should be deployed at each layer of the cyber hierarchy \cite{guo2015}. Honeypots are suggested by \cite{wang2017} and can absorb attack impact. Saturation attacks can also be prevented by minimising the number of device and server interfaces with remote access. For the disruption caused by ongoing DoS attacks, Li et al. \cite{li2019} suggested the use of predictive algorithms and historical data to estimate correct values and maintain grid stability. Meanwhile, SCADA and other control layer protocols need to introduce stronger authentication and improved security.

\subsection{Network Exploit Scenarios (\textit{A.2})}

Wireless ad-hoc network architectures are designed to be self-forming so that nearby nodes can organise themselves to define routes for data transmission. As stated previously, mesh networks may be deployed in the AMI (and possibly within the wider distribution and transmission domains for field sensors). However, the protocols used may be exploited to prevent these networks from forming and/or stabilising. This possibility was explored by \cite{yi2016}, who defined the puppet attack against the DSR protocol. Other ad-hoc routing protocols like RPL (Routing Protocol for Low-Power) or OLSR (Optimized Link State Routing) may also be vulnerable to similar attacks. The result of an \textit{A.2} scenario in the AMI may therefore be characterised as a customer domain network which has failed to converge and so no smart meter data can be delivered. Jamming attacks can similarly disrupt the normal operation of wireless network protocols \cite{chatfield2018}.

As suggested in \cite{pedramnia2018}, LTE-based cellular networks may be used as an AMI architecture. In this type of setup, bearers are created to link devices to the data network. Bearer assignment may be exploited in a method similar in concept to TCP SYN; channels are opened to the target and then not used. As a result legitimate access is denied \cite{pedramnia2018}. Once again, this would lead to the denial of AMI services.

Another possible \textit{A.2} scenario is the exploitation of AMI NAT systems, as identified by \cite{pedramnia2018}. Despite the introduction of IPv6, IPv4 is still widely used. This means that NAT is required for both translating between private and public IPv4 addresses, and for IPv4-to-IPv6 mappings \cite{pedramnia2018} \cite{nat64}. For example, if field sensors using IPv6 attempt to send data to core networks still operating on IPv4, the corruption of NAT64 mappings would deny such transmissions. In cases such as this and in LTE networks, \textit{A.2} scenarios will be characterised by a lack of connectivity between endpoints. We did not identify \textit{A.2} scenarios for the other sub-systems during our survey and suggest this as an area for further study.

To avoid network disruption, routing protocols should be secure against tampering attempts, and robust enough to re-converge efficiently. Such an approach was proposed by \cite{srikantha2015} whose topology configuration scheme is designed to maintain routing between PMUs and RNs. Similarly, \cite{yi2016} suggested that corrupted nodes be identified by their abnormal communications and isolated, given that WSN nodes depend on their neighbours for their network connectivity. For field networks, physical security is needed to prevent illegitimate devices from joining ad-hoc networks. Anomaly detection applied to network exploitation needs to monitor protocol activity rather than general communication traffic. For SMS links, \cite{pedramnia2018} cited the use of machine learning with bearer-related data to achieve this. They also suggest that migrating more widely to IPv6 could reduce the need for NAT mappings, thus reducing the risk of this type of DoS \cite{pedramnia2018}.

\subsection{Device Exhaustion Scenarios (\textit{B.1})}

Certain devices may be specifically targeted. An example in the Phasor Control sub-system is the relay nodes between PMUs and control systems, as highlighted by \cite{srikantha2015}. Given the large number of data collection points, relay nodes are used to ensure end-to-end delivery, sometimes also acting as aggregators. A targeted DoS attack on these devices can therefore disrupt the whole sub-system's communication flows, again leading to inaccurate measurements being collected. Furthermore, it is plausible to assume that relay node services in other sub-systems, such as the AMI, Load Frequency Controls, and Telemetry Controls, can be denied in this way too.

PLCs (and their interfaces) may also be actively targeted, as highlighted by \cite{yilmaz2018}. Like relay nodes, PLCs have a central role in data collection and aggregation, but are more closely integrated with core control systems. Therefore, deliberate disruption of PLCs will significantly impact management decisions. The same scenario may be applicable to RTUs too. However, devices such as these, which are found deeper within the smart grid architecture, should theoretically be more difficult to access and would require insider access or skilled adversaries. Therefore, \textit{B.1} scenarios in such systems will be characterised by the likely presence of skilled attackers or malicious insiders, and the hindering of data delivery to the core.

Finally, direct attacks within the AMI may be directed at smart meters, overwhelming their processing capacity as demonstrated by \cite{wang2017} and \cite{sgouras2014}. Unlike attacks on the AMI channels, forcing smart meters offline has the benefit of preventing both their communications with appliances and with head nodes directing traffic to the core. The location of smart meters within customer premises also exposes them as easier targets. Furthermore, \cite{wang2017} and \cite{sgouras2014} both demonstrated that AMI aggregator nodes and utility servers can be directly targeted. Attacks may be in the form of TCP SYN flood, HTTP flood, or other similar device-focused TCP/IP methods. Therefore, these \textit{B.1} scenarios will be characterised as disabled devices within the AMI tree, with diminished connectivity for all the nodes served by them.

\textit{B.1} is the device-focused equivalent of \textit{A.1}, so similar defensive measures may be applied. The aim is to knock particular devices offline, so suspicious traffic directed at those devices can be identified using anomaly detection. Anomaly profiles should be derived from the normal activity of those devices, with consideration of typical performance values, and critical nodes may be prioritised to reduce the overheads introduced by this. Predictive algorithms may again be applied to control systems to reduce the impact of lost telemetry data \cite{li2019}. To deal with possible insider threats, \cite{yilmaz2018} recommends strict monitoring and regulation of user privileges and activities.

\subsection{Device Compromise Scenarios (\textit{B.2})}

A possible compromise scenario is where smart grid devices are recruited into botnets. Whilst this is less probable in the deeper areas of the grid (given that those devices may be running proprietary software and may be difficult to physically access), it should not be completely disregarded as a possibility where malicious insiders are considered. Bot compromise is most likely to occur within edge systems (like the AMI), where some devices sit within customer premises and may be exposed to public networks. As highlighted in \cite{guo2015}, a population of smart meters (with similar manufacturers and models) could see the spread of bot code via an automated malware. IoT-based botnets (made up of smart appliances) are also a factor. Whilst a botnet compromise is itself not necessarily a DoS scenario, it provides a platform from which to launch such attacks. Additionally, bots provide backdoor access and can serve as vectors for smaller and more targeted DoS campaigns. Botnet-based DDoS, where smart meters are compromised, was studied by \cite{guo2015}, and similar botnet scenarios were considered by \cite{wang2017}, \cite{sgouras2017} and \cite{asri2015}.

Another \textit{B.2} scenario, primarily targeting the field sensors, is the forced depletion of physical resources on the target device. This is especially problematic in low power networks such as those in WSNs. For example, the Denial-of-Sleep attack defined by \cite{hoffmann2019} is designed to drain the energy of battery-powered sensors and control devices to render them non-functional. Whilst the impact of such an attack may be slower to manifest, it could have a longer-term impact on service quality as depleted nodes would need to be physically changed. The loss of nodes in the AMI would reduce its overall accuracy and functionality, but Denial-of-Sleep could plausibly be used against any type of battery-powered field device. \textit{B.1} scenarios may therefore be characterised by the presence of grid devices over which the operators do not have full control.

To stop nodes from being recruited into a botnet, devices must be regularly updated and patched with the latest software to minimise vulnerabilities. This can be challenging for difficult-to-access field sensors with limited capacity, which highlights the need for security-by-design in such devices. Login credentials must also be properly configured, as default passwords can be used to gain access \cite{antonakakis2017}. Battery depletion can be mitigated by designing more energy-efficient devices, and connecting nodes to the main power where possible. To deal with Denial-of-Sleep, \cite{hoffmann2019} suggested the need for an additional security layer for key exchanges to prevent meters being forced into FAC (Frequent Access Cycle).

Fig. 3 summarises and depicts the smart grid DoS scenario classifications described here, and Fig. 4 illustrates where DoS attacks sources may sit in relation to the smart grid.

\begin{figure*}[!t]
\label{fig3: attack classification}
\centering
\includegraphics[scale=0.55]{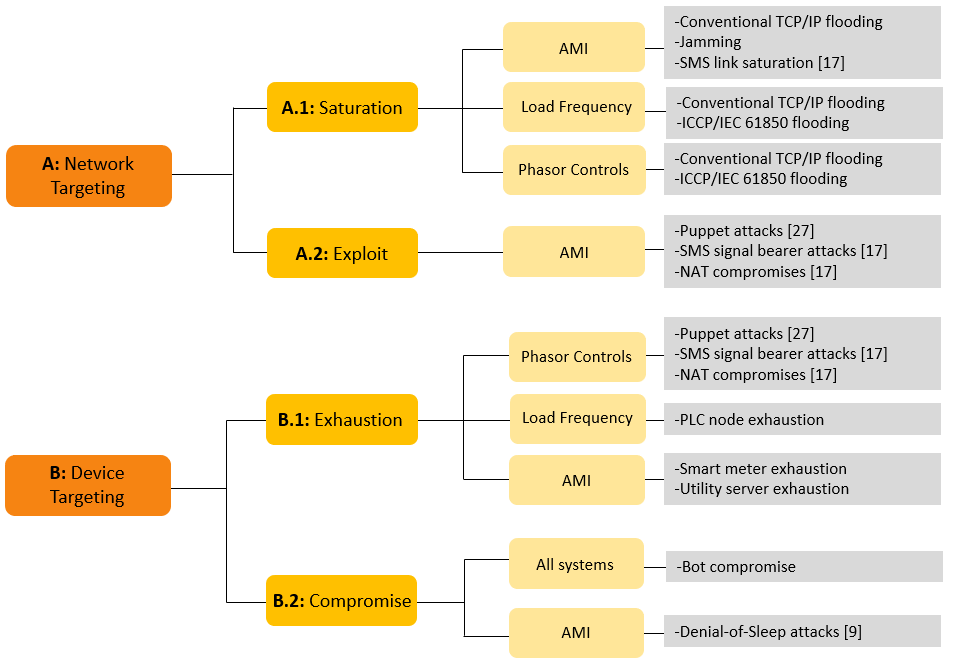}
\caption{Smart grid DoS attack classification tree.}
\end{figure*}

\begin{figure*}[!t]
\label{fig4: attack sources}
\centering
\includegraphics[scale=0.3]{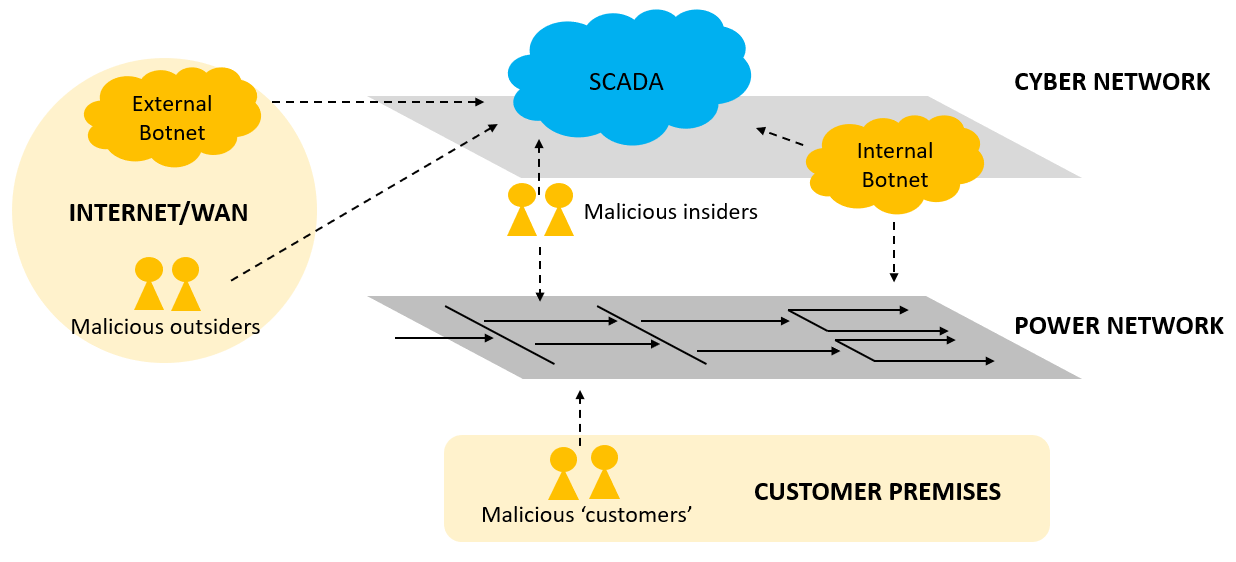}
\caption{Smart grid DoS attack sources.}
\end{figure*}

\section{Discussion}

In this work, we sought to uncover where and how DoS attacks may manifest in the smart grid. Given that this is still a relatively new technology, there have thankfully been very few real-world attacks thus far, for which (to our knowledge) detailed data is not available. Therefore, we turned to the available cyber-security literature and our understanding of smart grid architectures (combined with what we know of typical DoS attacks) for answers. We believe that this proved to be a sound methodology, as through the survey, we were able to enumerate and identify the sub-systems which could be targeted and how. This provides defenders with a means to prioritise their methods. It also provides future researchers with a reference point for what has been done so that other sub-systems may be considered as well. For example, areas for future DoS research may include attacks against generators and distributed power generation, and those originating from within the corporate networks of service provides or the Markets domain.

Through the survey, we were able to identify that the research community considers the AMI to be the most likely target of DoS attacks, with botnets being the most likely source. This suggests the need for better AMI security measures, and relates to other open research topics such as the security of IoT devices and WSNs. Meanwhile, the definitions of Denial-of-Sleep by \cite{hoffmann2019} and  puppet attacks by \cite{yi2016} demonstrate how the particularities of the smart grid can create space for new attack scenarios. Both scenarios were set in the AMI, once again highlighting the need for resilient security channels, as well as security-by-design in IoT technologies. It is also worth noting that these attacks may be feasible wherever mesh networks or battery-powered devices are used across the grid domains.

The potential vulnerability of other systems, despite their locations deeper within the grid network, is also apparent in the works of \cite{wei2012} and \cite{srikantha2015} who observed the impact of DoS on PMUs, and \cite{yilmaz2018} and \cite{liu2013} who did the same for PLCs and LFCs, respectively. These components are likely harder to impact remotely and without specialist knowledge, but may still be targeted by skilled attackers (e.g. those working on behalf of nation states). The location of such components, higher up within the grid hierarchy, also means that the impact of any successful attack will be felt throughout the grid. Therefore, these systems must be secured appropriately.

Furthermore, we believe that the proposed classification provides a novel perspective for smart grid DoS. There are many classifications of DoS attacks in the existing literature, but most focus on the methodology. This is appropriate for the high-level characterisation of generic DoS attacks, but we argue that in the context of smart grids, it is beneficial to characterise attacks by target and impact, as this helps to align them with the IT and OT layers. Simply, this view enumerates the grid's vulnerable points. Furthermore, it makes a distinction between attacks against the channel and against the device. This is not always clear in exploit-based DoS scenarios, but is significant in smart grids because the domain of attack and the domain of impact can be different. By determining where an attack is intended to cause damage, we can work towards expanding typical DoS defence methods.

Due to scope restrictions, the survey itself was not exhaustive and can also be expanded - the smart grid is a complex web of sub-systems spanning the IT, OT, and domains. Therefore, another possibility for future work is to build upon this survey to uncover more grid systems which may be vulnerable to DoS attacks, both known and novel. We assume that as smart grids become more well-established, the research into their security will also grow. Finally, as mentioned in Section 4.1, protocols developed for critical infrastructure may be further studied to uncover new DoS attack methods and possible exploits.

\section{Related Work}

The novelty of this work comes from our singular focus on DoS scenarios for smart grids, with special consideration for smart grid sub-systems, supported by a survey of the existing literature on the subject. Whilst there are many works considering the different types of DoS attack and the cyber-security challenges faced by smart grids, few works focus on both at the same time. In addition to this, most existing classification schemes deal with methodology whilst this work aims to highlight the relationship between attack targets and attack impact.

Huseinovic et al. \cite{huseinovic2018} developed a taxonomy of possible DoS attacks based on available literature, followed by a discussion of defensive strategies. Similar to this work, they explored different taxonomy perspectives, including one that considers which grid applications are targeted. However, they put more emphasis on analysing the security measures against each attack type and did not provide in-depth characterisations as we have. Otuoze et al. \cite{otuoze2018} also considered an alternative classification perspective by looking at threat sources, identifying both technical (i.e. infrastructural, operational, data) and non-technical (environmental, policy) sources, but did not provide details on DoS scenarios.

Wang and Lu \cite{wang2013} contributed a thorough survey of cyber-security challenges in the smart grid, with a section dedicated to DoS threats. However, they did not examine different sub-systems to characterise and classify attacks as we have. El Mrabet et al. \cite{el2018cyber} conducted a similar survey but also did not offer characterisations. In their extensive cyber-security survey, Tan et al. \cite{tan2017survey} did consider different smart grid components, but organised them differently as sub-systems (AMI, SCADA/WAMS) and data-generating devices (smart meters, PMUs, etc). Meanwhile, they did not look in detail at DoS scenarios. Our work sits alongside this to provide an alternative perspective.

Lastly, Ramanauskaite and Cenys \cite{ramanauskaite2011} created a detailed taxonomy of DoS attack types and the defences against them, including considerations of attack source, exploited vulnerabilities, method, target type, and rate. However, smart grids were not in the scope of their work and so they did not consider the grid-specific attack targets. As with \cite{tan2017survey}, we believe our work sits alongside this to help focus in on smart grid-specific DoS threats.

\section{Conclusions}

Smart grids are designed to make the generation and provision of power services more efficient and sustainable through the integration of IT technologies. However, this exposes the OT to cyber threats as seen in conventional networks and the Internet. DoS and DDoS attacks are among the most prevalent of these threats. This work provides a survey and a summary of the grid sub-systems that may be targeted, and characterises several possible DoS scenarios, alongside a target-based classification scheme to support this new perspective. Overall, we hope to have highlighted areas of smart grid vulnerability and set a foundation for better DoS defence and mitigation.

\section*{Acknowledgment}

This work is funded by and a part of Energy Shield, a project under the European Union's H2020 Research and Innovation Programme.

\label{a: bibliography}
\bibliographystyle{splncs04}
\bibliography{bibliography}

\end{document}